\documentstyle[12pt,epsfig]{article}

\begin{document}

\title{Charm production by cosmic muons}

\author{Francesco Vissani\footnote{e-mail: vissani@lngs.infn.it},\\
Laboratori Nazionali del Gran Sasso, INFN, Theory Group\\
Strada Statale 17/bis, km 18+910,\\ 
I-67010, Assergi (L'Aquila) Italy}
\date{}
\maketitle

\begin{abstract}
Narrow muon bundles 
in underground detectors permit to study 
muoproduction reactions that take 
place in the surrounding rock.
We analyze the relevance of 
a QED+QCD reaction,
muoproduction of ``open charm''.
The contribution to double muon 
events is estimated to be 
$4-8$\% of the one  
due to QED ``trident'' process, 
for an ideal detector 
located under a rock depth of 
3 km water equivalent, and 
an observation threshold 
of 1 GeV.
\end{abstract}

In recent years, there has been a certain 
experimental \cite{MACRO,LVD}
and theoretical interest \cite{olga,BS,kelner,music2}
on  ``narrow muon bundles'' 
(multiple muons with a lateral 
separation less than a few meters)
in underground detectors.
These events have been observed 
as a peak at small lateral separation,
and interpreted as an induced flux.
In fact, the energetic muons 
that propagate underground 
in roughly $\sim 1$\% of the cases 
interact and produce other muons.
Thence, an analysis of these 
events requires to consider high energy
muoproduction processes, in the rock 
surrounding the detector.

Up to now, the process considered was the 
muon ``trident'' reaction \cite{trident}
$\mu\, Z\to \mu\, Z\, \mu^+\mu^-$, where a muon pair 
is formed  in the field of the 
nucleus\footnote{It is assumed that 
an effective rejection of muoproduced 
$\pi^\pm,\ \gamma,$ e$^\pm$ ...\ 
can be achieved.}.
For muons propagating in high $Z$ materials 
an amplification factor $Z^2/A$ 
($= 5.5$ for standard rock, $A$=22 and $Z$=11) 
is present, due to coherent  character of the reaction.
This interpretation has been pursued 
since the first evidences obtained 
in cosmic ray experiments \cite{Morris-Stenerson}.
The trident reaction leads mostly to 
narrow bundles of three or two muons in 
an underground detector (one produced 
muon may stop before reaching the detector);
a reference ratio in an ideal detector
is of $3$ double muons {\em per}
triple muon, for a threshold of observation
$E_h = 1$ GeV, and a depth $h=3$ km w.e.\ 
of standard rock.
Recent studies \cite{kelner,music2}, however, 
suggest that existing interpretations are 
insufficient to quantitatively account for 
the whole ``narrow muon'' data set. 

In this work we analyze the 
role of another high energy process
as source of  prompt muons:
production of charmed states due to 
cosmic (atmospheric) muons, 
whose relevant energies range from\footnote{We 
neglect the energy loss in the rock of the charmed hadrons $X_c,$
for a 200 GeV $D^\pm$ meson  travels on average just 
3 cm in the rock before decaying.}  
$E\sim 100$ GeV up to tenths of TeV`s
(for studies in laboratory, see \cite{muoprod}).
More specifically, we 
are concerned with the
``open charm'' reaction of muoproduction: 
$\mu\, N\to \mu\, c\bar{c} X$
($X$ denotes a byproduct which 
does not concern us). 
This process  is stipulated by
QED and QCD interactions, while
weak interactions provide the  
instability of charmed states: 
$c\to X_c\to \mu X.$

In order to obtain a simple estimate 
of the flux of double muons due to this 
process, we adopted the collinear 
approximation, considering only how the 
initial muon with energy $E$ branches 
into those of the final muons ($E'$ and $E''$) 
and proceeded in the following 
way:

{\em 1)}\ \  First, we calculated the 
cross section of muoproduction
$d \sigma_{\mu N\to \mu c\bar{c}X}/dE'dE_c $
at leading order (LO)
in $\alpha_s,$
double differential 
in the energies of the scattered muon, 
$E'$ and of the charm, $E_c$ (see appendix).
This can be done with a
limited effort by following 
the calculations documented 
in \cite{lw}, where the cross section integrated 
over the hadronic phase 
space $d\Phi_{hadr}$ was obtained:
In fact, neglecting the gluon mass, the 
differential expression is simply
$d \Phi_{hadr}=dE_c/(8\pi E_\gamma),$ 
where $E_\gamma=E-E'$ is the energy 
of the virtual photon emitted by the muon\footnote{Also, 
we found convenient to relate $E_c$ to the zenith angle 
and velocity of emission in the gluon-gamma c.m.s.\ as follows:
$E_c/E_\gamma=(1+\beta_c^* \cos\theta_c^*)/2,$
where $\beta_c^*=\sqrt{1-4 m_c^2/(p+q)^2}$ 
($p$ and $q$ are the momenta of the 
gluon and of the virtual photon)}.
In the actual calculation, 
that requires integrating over 
the photon virtuality $Q^2$ and the 
gluon momentum fraction $x,$
we use the GRV98 gluon distribution \cite{grv},
and set: $m_c=1.5,$ 1.35 or 1.2 GeV. 
We  multiplied the cross section by 
the factor $K(E)=\sigma_{NLO}/\sigma_{LO}$
(where $\sigma$ is the total cross section) 
to describe next-to leading order
QCD effects \cite{NLO,smith,brian}. 
The differential cross section increases 
with $E'$ with a ``1/$v^2$ behavior''
and than rapidly  decreases to zero  
in the range of energies of interest;
instead, it is rather mildly distributed in  $E_c.$ 
The total cross section $\sigma$ 
increases with $E,$  due to the smaller values of $x$ 
that are probed by the virtual photon, and 
to well known characteristics 
of the gluon distribution.
Its value is $4 \times 10^{-32}$ cm$^2$ 
when $E=1$ TeV for $m_c=1.35$ GeV
(almost equal to the trident cross section 
at the same energy); LO cross section 
increases by 50\% if $m_c$ is lowered to 1.2 GeV, 
while decreases by 30\% if $m_c$ is 1.5 GeV.

{\em 2)}\ \  We estimate a ``scaling''
probability $dP_{c\to \mu}/dw\equiv 
BR_{c\to \mu}\times \rho_{c\to\mu}(w)$ 
that a charm yields
a muon with a certain energy fraction $w=E''/E_c,$
by first hadronizing the charm into a $D$ meson
(using the normalized distribution of \cite{Peterson} 
with $\epsilon_D=0.135$)
and then letting it decay  with a 
$K_{\mu 3}$ distribution (that is, 
retaining only the $D$ mass, 
and neglecting the $Q^2$ dependence
of the form factors).
The resulting normalized probability $\rho_{c\to\mu}(w)$ 
falls strongly with the 
energy fraction $w;$
the median of the distribution is in fact 
$\langle E'' \rangle = 0.15 \times E_c .$
We took as an effective 
branching ratio of charm 
into muons the value
$BR_{c\to \mu}=8$\% \cite{PDG}, 
and multiplied the result by 
two, to account for the fact that 
a charm or an anticharm can 
yield a muon\footnote{Existing 
underground detectors do not 
distinguish between ``same charge'' 
and ``opposite charge'' double muon events.}.
Notice, incidentally, that 
the corresponding yield 
of triple muon is negligible, 
due to an {\em a priori} 
4\% suppression factor.

{\em 3)}\ \  At this stage, we can calculate 
the cross section $d \sigma_{\mu N\to \mu\mu}/dE'dE'',$
where $E''$ is the energy of the produced muon,
and, with that, the cross section\footnote{We 
consider only those events whose vertex is  
{\em not} contained in the detector. 
Those events profit of a large 
effective target mass, and correspond, 
in a sense, to the celebrated neutrino-induced 
single muon signal.} 
$\sigma_{\mu N\to \mu\mu}(E, f)$ for production
of two muons, each with a fraction of 
the initial muon energy greater than $f.$  
Due to the behaviors of 
$d\sigma_{\mu N\to \mu c\bar{c} X}$  and
$d P_{c\to \mu}$ with $E'$ and $E''$
mentioned above, this cross section 
diminishes dramatically with $f;$
when $E=1$ TeV, it drops down by one order 
of magnitude already when $f\approx 0.07.$ 
This cross section enters the elementary
yield of double muons in the detector, 
which depends linearly on the 
infinitesimal depth crossed $dh'$ (in gr/cm$^2$):
\begin{equation}
dY_{\mu\to \mu\mu}(E,{h'})= 
dh'\times  N_A\times \sigma_{\mu N\to \mu\mu}(E,f) \ 
\mbox{ where } f=\frac{E_{h'}}{E}
\label{yield}
\end{equation}
$N_A=6.023\times 10^{23}$ 
is the number of nucleons in 1 mole 
(multiplying by $dh',$ we obtain the
density of targets {\em per} cm$^2$).
The energy losses are evaluated 
in continuous energy loss approximation, 
$E_{h'}=(E_h+\epsilon) 
\exp[\, (h-h')\, /\, h_0\, ] -\epsilon,$ where 
$\epsilon\approx 600$ GeV and $h_0\approx 2.5$ km w.e.\ 
are phenomenological parameters, and $E_h=1$ GeV is the  
(typical) threshold for detection.
Multiplying this by the single 
muon flux differential in $dE$, 
${d F_\mu},$ we get the differential 
double muon flux induced 
by ``open charm'' reaction
at the depth $h.$
The total flux is then:
\begin{equation}
{F_{\mu\mu}}(h)
=\int_0^h 
\int_{2 E_{h'}}^\infty 
{d F_\mu}(E,h')
\times
dY_{\mu\to \mu\mu}(E,{h'})
\label{mu3flux}
\end{equation}
where we integrated over the depth of production $h' ,$
and the energy $E$ of the primary muon at this 
depth (namely, where the reaction takes place);
$E$ was related to the energy at the surface $E_0$ 
in the continuous energy loss approximation, 
which permits us to evaluate ${d F_\mu}(E,h')/dE$ by using
the (approximate) expression for the flux at 
the surface given in \cite{Gaisser}
(namely, ${d F_\mu}/dE_0=0.14\times 
E_0^{-2.7}\times ...$ /(cm$^{2}$ s sr GeV)).
The results are shown in the figure,
for muons arriving from 
the vertical direction.
\begin{figure}[thb]
\leftline{\epsfig{file=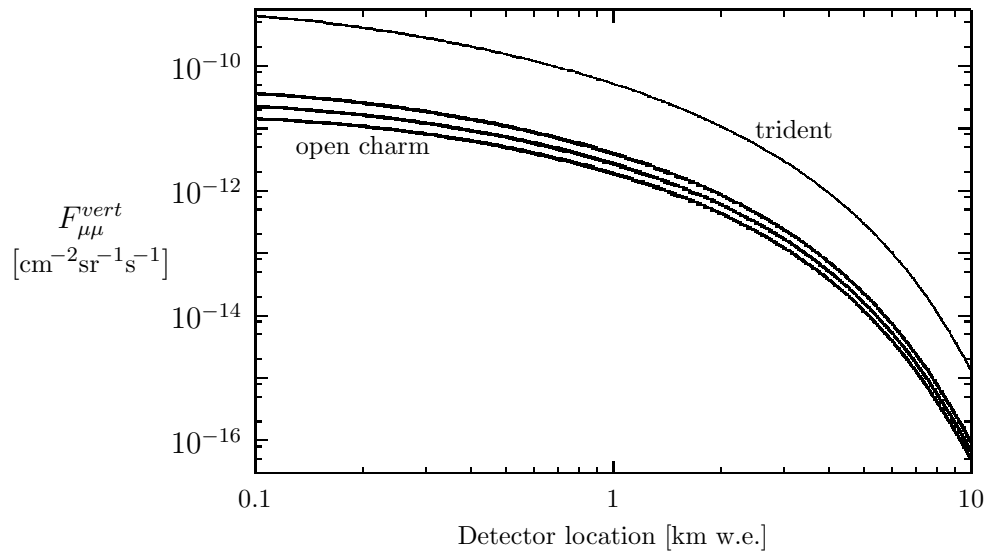,width=13cm}}
\caption{Flux of narrow double muons 
due to open charm formation (thick curves, 
for $m_c=1.5,1.35,1.2$ GeV from lower to higher one) 
and to the trident reaction (thin curve,
calculated with the cross section given in \cite{kelner}).} 
\label{fig:}
\end{figure}

The contribution of open charm reaction is 
not large; for instance, at a depth 
of 3 km w.e.\ it is just $4-8$\% of the 
one due to the trident process. 
Equivalently, it can be compared with the flux of single muon: 
we get $F_{\mu\mu}/F_\mu=0.7,1,1.4\times 10^{-5}$ 
in the case of open charm reaction (for $m_c=1.5,1.35,1.2$ GeV)
while $F_{\mu\mu}/F_\mu=1.8\times 10^{-4}$ in the case of trident reaction.
For an ideal detector, it
would require to accumulate several hundredth
(trident narrow double muon) events, 
to become statistically interesting.
The smallness of the result 
has to be attributed to the relatively small
value of $BR_{c\to \mu},$ and to 
the effective leakage of energy 
of the virtual photon, during 
the conversion $\gamma^*\to c\to D\to \mu$
(while for tridents, 
$\gamma^*$ immediately materializes into muons).
However, this contribution 
is not negligible if 
one aims at reaching the precision of
$5-10$\% in the predictions.

The following remarks 
illustrate other aspects 
of this result:\\ 
(a) going to shallower depths,  
the double muon flux increases, 
though less rapidly than the 
single muon flux: in fact,
the  effective target increases with the depth
(but, of course, also the background increases);\\
(b) conversely, in deeper sites the relative 
contribution of the open charm process 
becomes more important,
due to more energetic primaries--$E$ increases
(but there are practical limitations, 
due to the time of data taking 
and area of the installation); \\
(c) keeping the depth fixed, and changing the 
angle of observation, there is an increase 
of $F_{\mu\mu}$ moving toward the horizon, directly
related to the increase of $F_\mu$ 
(but the actual geometry of the rock 
in the underground site has an 
essential role in practical considerations);\\
(d) in water or ice, the trident curve 
would reduce by $ \sim 1.5$ in comparison with the 
open charm one, due to the $Z^2/A  $ factor.
This would somehow emphasize the open charm contribution
(but it should be reminded that no plan exists 
to have an underwater detector, 
with large area and capable to achieve a 
good discrimination at small lateral distances).

In conclusion, it seems to us 
rather difficult to account for a large fraction
of narrow muon bundles on the basis of the 
open charm process. Thus, the
chances of studying heavy quark physics 
with existing underground (or underwater)
detectors are quite limited. 
This result, however, adds motivations 
for further search of unexpected sources of background
(and, possibly, new sources of prompt muons) when we 
recall the difficulties to understand existing 
data on narrow muon bundles.
For the future experimental 
perspectives, we consider interesting 
the possibility to achieve energy and charge 
identification of the muons in  
underground detectors, as a possible handle to 
separate various components in a 
narrow muon bundles data set (for instance, we found 
that the average energy of the parent muon--that 
forms two muons through the open charm process--is 
rather large, above 1 TeV). 
However, even if it will be possible to obtain  
sufficient statistics and control of the systematics, 
an attempt to proceed further and 
extract a signal of production 
of heavy quarks from studies of 
narrow muon bundles
will need more refined calculations:
To accurately describe the 
NLO effects \cite{smith}, hadronization, 
and decay of charmed states \cite{bugaev,semilep};
but also those effect in the muon propagation, 
that are necessary to model the 
lateral distribution of the events in the 
underground detectors \cite{paolo,music,music2}.
In fact, the relatively large transverse 
momenta $p_\perp \sim m_c$ that result from 
charm production and decay lead to 
larger lateral separations than 
those due to the trident reaction, 
and this could be of interest to characterize 
the charm induced events.

\vskip.5cm
\noindent{\Large \bf Acknowledgments}
\vskip.3cm
\noindent I would like to 
express my gratitude to several people:
to V.S.\ Berezinsky for guide and support,
and for informing me on 
the work \cite{v}, that the 
present study followed 
rather closely;
to E.\ Scapparone for pointing 
my  interest on the physics of 
``narrow muon bundles''; 
to G.\ Battistoni, 
S.\ Cecchini, 
R.P.\ Kokoulin,
V.A.\ Kudryavtsev,
P.\ Lipari and
O.G.\ Ryazhskaya
for helpful discussions 
on cosmic muons; 
A.V.\ Butkevich for explaining me
how to include the nuclear effects;
to B.W.\ Harris for help with 
NLO charm production cross section.

\newpage
\noindent{\Large \bf Appendix: Formul\ae\ for LO cross sections}
\vskip.3cm
The LO cross section, differential
in $y=E_\gamma/E$ and $z=E_c/E_\gamma$  is:
\begin{equation}
\begin{array}{l}
\displaystyle
\frac{d\sigma_{\mu N\to \mu c\bar{c} X}}{dy dz}=
\frac{\alpha^2}{9 M E} 
\  \int d\log{Q^2} \ \int d\log{x} \ 
  \alpha_s(\mu^2)\ G(x,\mu^2)  \\[2ex] \displaystyle \ \ \ \ \ \ \ 
\left[
 \left(1-\frac{2 m_\mu^2}{Q^2}+2\ g(y)\right) \frac{df_T}{dz} + 
 \left(1-\frac{2 m_\mu^2}{Q^2}+6\ g(y)\right) \frac{df_L}{dz} \right]
\label{loxsec}
\end{array}
\end{equation}
where $M=0.938$ GeV, $m_\mu=0.106$ GeV 
and $\alpha=1/137;$ $\alpha_s$ is the strong  coupling,
$G$ the gluon distribution function, 
and $\mu^2$ the factorization scale
($\mu^2=4 m_c^2+Q^2$ in present calculation).
In order to describe nuclear effects in ``standard rock'' nuclei,
we followed \cite{smirnov}, and 
weighted the gluon distribution function by the 
density: 
$r^{s.r.}(x) \approx 1.26 \times x^{0.073}\times (1-0.3\, x);$ 
see eq.\ 1 and fig.\ 2 in that work\footnote{Notice 
that here $x$ is the momentum fraction of the gluon 
(the particle that feels the nuclear effects),
as contrasted with $x_F\equiv Q^2/(2 Pq),$ the variable 
introduced to parameterize the structure 
function $F^{c\bar{c}}(x_F,Q^2).$}.
The adimensional functions
introduced in eq.\ \ref{loxsec}
are $g(y)=(1-y)/y^2$ and
\begin{equation}
\begin{array}{l}
\frac{df_T}{dz} =g_1-g_0 + x_c g_1-\frac{x_c^2}{4}  (g_2+2 g_1)-
2 x_Q (g_1-\frac{x_c}{4} g_2) + 2 x_Q^2 g_1 \\[1ex]
\frac{df_L}{dz} =2 x_Q (g_0-\frac{x_c}{2} g_1) - 2 x_Q^2 g_0
\end{array}
\end{equation}
where $g_n=1/z^{n}+1/(1-z)^{n}$ for $n=0,1,2;$
$x_c=2 m_c^2/pq$ and $x_Q=Q^2/(2pq)$ 
are bound by $x_{min}$ to be 
lower than unity. 
The non-trivial limits are: 
$(i)$ $x_{min}=({ m_c^2\, g_1(z)+Q^2})/(2 M E_\gamma),$ which results
from the kinematics of the 
$\gamma^* g\to c\bar{c}$ process (setting the gluon mass to zero),
and $(ii)$ $Q^2_{min}\approx m_\mu^2/g(y)$ which results from
setting to zero the scattering angle in the laboratory frame;
the limits on $y$ and $z,$ and $Q^2_{max}$ 
follow by consistence.
Notice that the cross section can be easily integrated
analytically over $z,$ which amounts to replace:
\begin{equation}
g_0\to 2 \beta^*,\ \ \ 
g_1\to 2 \log\frac{1+\beta^*}{1-\beta^*},\ \ \ 
g_2\to \frac{8 \beta^*}{1-(\beta^*)^2}
\end{equation}
and, also, $g_1(z)\to g_1(1/2)=4$ 
in the expression of $x_{min};$
the resulting expression for 
$d\sigma_{\mu N\to \mu c\bar{c} X}/dy$
is equivalent to  the one shown 
in the appendix of \cite{lw}.
The cross section that enters
the expression of the double muon 
yield (eq.\ \ref{yield}) is obtained as:
\begin{equation}
\sigma_{\mu N\to \mu\mu}(E,f)=
2\times BR_{c\to \mu}\times \int^{1-f}_f dy \int^1_{f/y} dz\  
\frac{d\sigma_{\mu N\to \mu c\bar{c} X}}{dydz} \times
\int_{f/(yz)}^1 dw\ \rho_{c\to \mu}(w)
\end{equation}
The last integral corresponds 
to an integral probability,
and can be tabulated separately to
simplify the calculation.


\begin{thebibliography}{99}
\bibitem{brian}
We use a $K(E)$ factor calculated by B.W.\ Harris,
for $m_c=1.5$ GeV and $Q^2>2$ GeV. 
Its approximate numerical expression is
$K\approx 1.5-\log(E/E_0)/3 +\log^2(E/E_0)/200,$ 
when $E<E_0=500$ GeV; above $E_0,$ 
the coefficient of the log halves, 
while the one of log$^2$ doubles.
\end{thebibliography}
\end{document}